# Plasmon modes of double-layer graphene at finite temperature

Dinh Van Tuan, Nguyen Quoc Khanh

*Department of Theoretical Physics, National University in Ho Chi Minh City, 227-Nguyen Van Cu Street, 5th District, Ho Chi Minh City, Vietnam*

___

**Abstract**

We calculate the dynamical dielectric function of doped double-layer graphene (DLG), made of two parallel graphene monolayers with carrier densities $n_1$, $n_2$, respectively, and an interlayer separation of $d$ at finite temperature. The results are used to find the dispersion of plasmon modes and loss functions of DLG for several interlayer separation and layer densities. We show that in the case of $n_2=0$, the temperature plasmon modes are dramatically different from the zero temperature ones.

___

## 1. Introduction

Graphene is a two-dimensional electron system that has attracted a great deal of attention because of its unique electron properties [1] and its potential as a new material for electronic technology [2, 3]. The main difference of 2D graphene compared with 2D semiconductor system is the electronic energy dispersion. In 2D semiconductor systems, the electron energy depends quadratically on the momentum, but in graphene, the dispersions of electron are linear near K, K' points of the Brillouin zone [4]. Because of the difference of electronic band structure, there are many graphene properties which are significantly different from the behavior of electrons in the ordinary 2D systems [5].

In this paper, we consider a DLG system formed by two parallel single-layer graphene (SLG) separated by a distance $d$. The DLG is fundamentally different from the well-studied bilayer graphene [6] because there is no inter-layer tunneling, only inter-layer Coulomb interaction. Spatially separated two-component DLG can be fabricated by folding an SLG over a high-insulating substrate [7].

Recently, Hwang and Das Sarma [8] have investigated the plasmon dispersion and loss function in doped DLG at zero temperature and found the surprising results. They have shown that the plasma modes of an interacting DLG system are completely different from the double-layer semiconductor quantum well plasmons. In the long-wavelength limit the density dependence of plasma frequency are given by $\left(\omega_0^+\right)^2 \propto \sqrt{n_1}+\sqrt{n_2}$ for optical plasmon and $\left(\omega_0^-\right)^2 \propto \sqrt{n_1 n_2}/\left(\sqrt{n_1}+\sqrt{n_2}\right)$ for acoustic plasmon compared to $\left(\omega_0^+\right)^2 \propto N$ and $\left(\omega_0^-\right)^2 \propto n_1 n_2 / N$ in ordinary 2D systems, where $N = n_1+n_2$.

In this paper we investigate the temperature effects on the plasmon mode and loss function of DLG for several interlayer separations and layer densities.

## 2. Theory

In graphene, the low-energy band Hamilton is well-approximated by a two dimensional Dirac equation for massless particles, the so-called Dirac-Weyl equation [9],

$$H = \hbar v_F \left(\sigma_x k_x + \sigma_y k_y\right) \qquad (1)$$

where $v_F$ is the 2D Fermi velocity, $\sigma_x$ and $\sigma_y$ are Pauli spinors and **k** is the momentum relative to the Dirac points. The energy of graphene for 2D wave vector **k** is given by:

$$\varepsilon_{\mathbf{k},s} = s v_F |\mathbf{k}| \qquad (2)$$

where $s = \pm 1$ indicate the conduction (+1) and valance (-1) bands, respectively, and $v_F$ is the Fermi velocity of graphene and $\hbar = 1$ throughout this paper. The density of states is given by $D(\varepsilon) = g|\varepsilon|/(2\pi v_F^2)$, where $g = g_s g_v = 4$ accounts for the spin ($g_s = 2$) and valley ($g_v = 2$) degeneracy. The Fermi momentum ($k_F$) and the Fermi energy ($E_F$) of 2D graphene are given by $k_F = \sqrt{4\pi n/g}$ and $E_F = v_F k_F$ where $n$ is the 2D carrier density.

In RPA, the dynamical dielectric function of SLG becomes

$$\varepsilon(q,\omega,T) = 1 - v_c(q)\Pi(q,\omega,T) \qquad (3)$$

where $v_c(q) = 2\pi e^2/\kappa q$ is the 2D Fourier transform of the Coulomb potential and $\Pi(q,\omega,T)$, the 2D polarizability at finite temperature, is given by the bare bubble diagram [10]

$$\Pi(q,\omega,T) = g \lim_{\eta \to 0^+} \sum_{s,s'=\pm} \int \frac{d^2\mathbf{k}}{(2\pi)^2} \frac{1 + ss'\cos(\theta_{\mathbf{k},\mathbf{k+q}})}{2} \frac{n_F(\varepsilon_{\mathbf{k},s}) - n_F(\varepsilon_{\mathbf{k+q},s'})}{\omega + \varepsilon_{\mathbf{k},s} - \varepsilon_{\mathbf{k+q},s'} + i\eta}. \qquad (4)$$

Here $n_F(\varepsilon) = \{\exp[\beta(\varepsilon - \mu_0)] + 1\}^{-1}$ is the Fermi-Dirac distribution function, $\mu_0 = \mu_0(T)$ being the noninteracting chemical potential determined by the conservation of the total electron density as

$$\frac{1}{2}\left(\frac{T_F}{T}\right)^2 = F_1(\beta\mu_0) - F_1(-\beta\mu_0) \qquad (5)$$

where $\beta = 1/k_B T$ and $F_n(x)$ is given by

$$F_n(x) = \int_0^\infty \frac{t^n dt}{1 + \exp(t - x)} \qquad (6)$$

The limiting forms of the chemical potential in low and high temperature are given by [11]

$$\mu_0(T) \approx E_F\left[1 - \frac{\pi^2}{6}\left(\frac{T}{T_F}\right)^2\right], \quad \frac{T}{T_F} \ll 1 \qquad (7)$$

$$\mu_0(T) \approx \frac{E_F}{4\ln 2}\frac{T_F}{T}, \quad \frac{T}{T_F} \gg 1 \qquad (8)$$

Recently, MacDonald and coworkers [12] have presented the following semi-analytical expressions for the imaginary $[\Im m\, \Pi(q,\omega,T)]$ and the real $[\Re e\, \Pi(q,\omega,T)]$ parts of the dynamical polarizability:

$$\Im m\, \Pi(q,\omega,T) = \frac{g}{4\pi}\sum_{\alpha=\pm}\left\{\Theta(v_F q - \omega)q^2 f(v_F q, \omega)\left[G_+^{(\alpha)}(q,\omega,T) - G_-^{(\alpha)}(q,\omega,T)\right]\right.$$
$$\left. + \Theta(\omega - v_F q)q^2 f(\omega, v_F q)\left[-\frac{\pi}{2}\delta_{\alpha,-} + H_+^{(\alpha)}(q,\omega,T)\right]\right\}, \qquad (9)$$

$$\Re e\, \Pi(q,\omega,T) = \frac{g}{4\pi}\sum_{\alpha=\pm}\left\{\frac{-2k_B T \ln\left[1 + e^{\alpha\mu_0/(k_B T)}\right]}{v_F^2} + \Theta(\omega - v_F q)q^2 f(\omega, v_F q)\right.$$
$$\left. \times \left[G_-^{(\alpha)}(q,\omega,T) - G_+^{(\alpha)}(q,\omega,T)\right] + \Theta(v_F q - \omega)q^2 f(v_F q, \omega)\left[-\frac{\pi}{2}\delta_{\alpha,-} + H_-^{(\alpha)}(q,\omega,T)\right]\right\} \qquad (10)$$

where

$$f(x, y) = \frac{1}{2\sqrt{x^2 - y^2}}, \qquad (11)$$

$$G_{\pm}^{(\alpha)}(q,\omega,T)=\int_{1}^{\infty}du\frac{\sqrt{u^{2}-1}}{\exp\left(\frac{|v_{F}qu\pm\omega|-2\alpha\mu_{0}}{2k_{B}T}\right)+1}, \qquad (12)$$

$$H_{\pm}^{(\alpha)}(q,\omega,T)=\int_{-1}^{1}du\frac{\sqrt{1-u^{2}}}{\exp\left(\frac{|v_{F}qu\pm\omega|-2\alpha\mu_{0}}{2k_{B}T}\right)+1}. \qquad (13)$$

The DLG dielectric function is obtained from the determinant of the generalized dielectric tensor and has the following form

$$\varepsilon_{do}(q,\omega,T)=\varepsilon_{1}(q,\omega,T)\varepsilon_{2}(q,\omega,T)-v_{12}(q)v_{21}(q)\Pi_{1}(q,\omega,T)\Pi_{2}(q,\omega,T) \qquad (14)$$

Here $\varepsilon_{1}(q,\omega,T)$ and $\varepsilon_{2}(q,\omega,T)$ are the dynamical dielectric functions of individual layers given by the Eq.(3), $v_{12}(q)=v_{21}(q)=2\pi e^{2}\exp(-qd)/(\kappa q)$, with $\kappa$ is the background lattice dielectric constant, are the interlayer Coulomb interaction matrix elements.

The spectrum of the collective excitations is obtained from the zeros of the real part and the imaginary part of the double-layer dielectric function describes the damping of collective modes.

### 3. Numerical results

In this section, we calculate the plasmon dispersion and loss function of DLG at zero and finite temperatures for different layer separations and densities.

#### 3.1. The plasmon dispersion

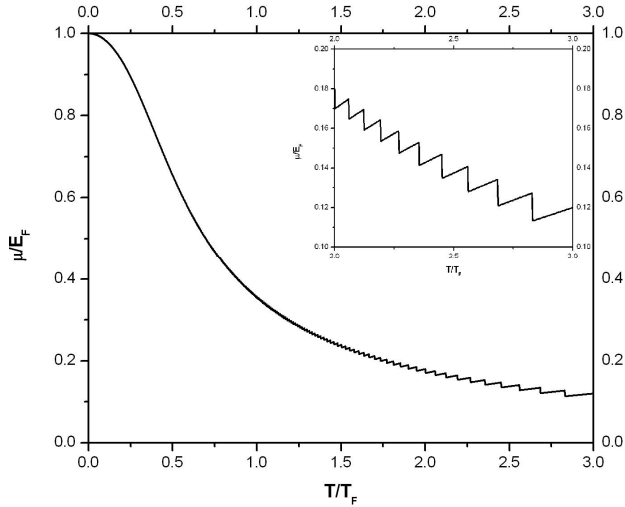

FIG. 1. The chemical potential of graphene

In Fig. 1, we show the calculated chemical potential which is used for calculating the dielectric function at finite temperature. In this Fig., the chemical potential show the strange behavior especially in the high temperature region.

In Fig. 2, we show the plasmon dispersions of DLG for several layer separations $d$ and temperature. As shown in Fig. 2, the frequency of the acoustic mode $\omega_{-}$ decreases compared with the SLG plasmon mode at the same temperature, while the optical mode $\omega_{+}$ shifts to higher energy. For low temperatures, the acoustic mode $\omega_{-}$ approaches the $\omega=v_{F}q$ line (dot-dashed line) in the high-energy region. More interestingly, when $T<T_{C}$ ($T_{C}\approx 0.6T_{F}$), the acoustic mode $\omega_{-}$ is below that at zero temperature, while for $T>T_{C}$, both acoustic and optical modes are above those at zero temperature. Furthermore, as $T$ increases ($T>T_{C}$) both acoustic and optical modes shift to higher energy. As $d$ increases both the acoustic mode $\omega_{-}$ and the optical mode $\omega_{+}$ in the low-frequency region approach the SLG plasmon at the same temperature. For large momentum ($q>k_{F}$), we find that the mode dispersion is unchanged when $d>100A^{0}$

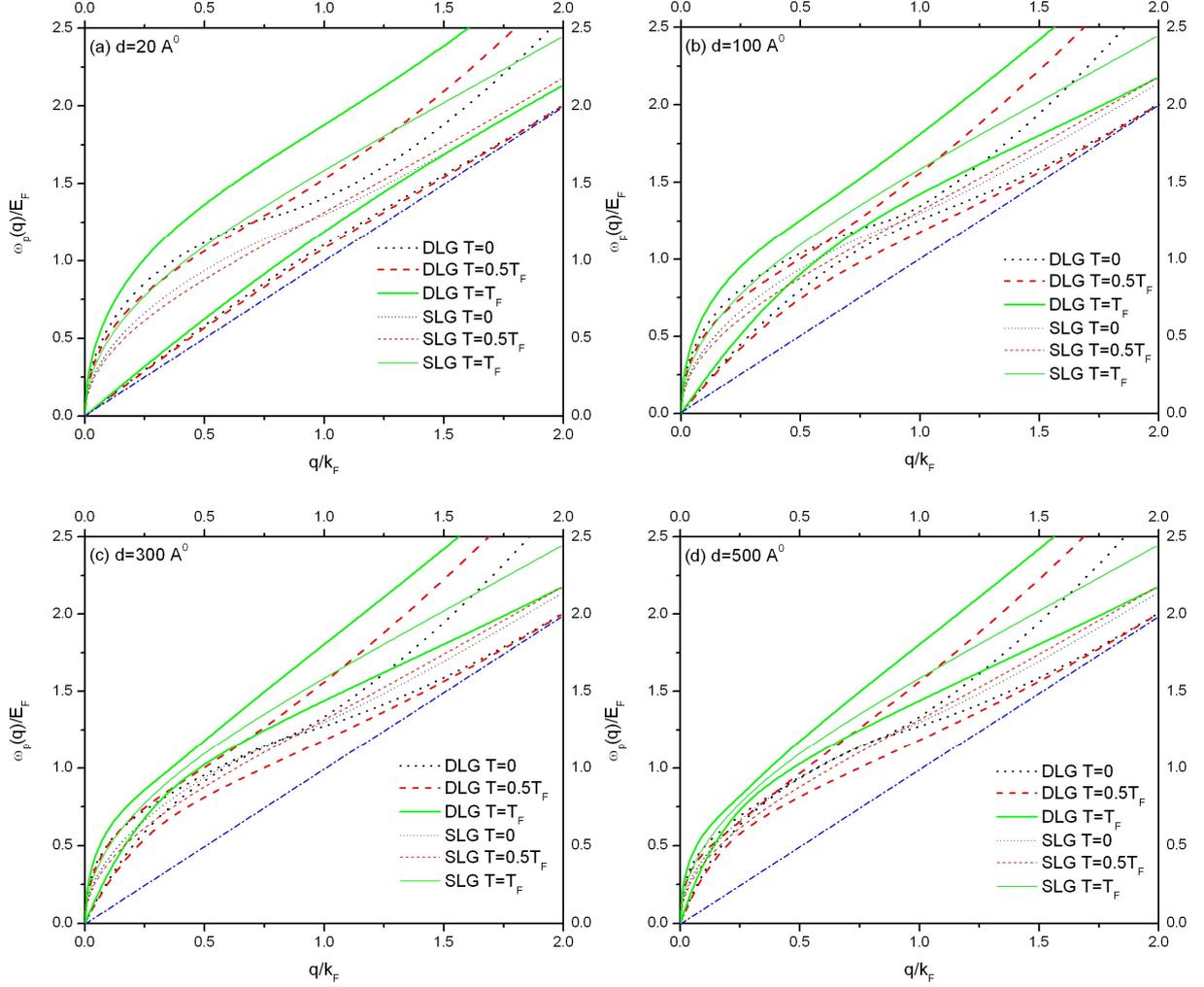

FIG. 2. (Color online) Plasmon dispersions of DLG for several layer separations at $T = 0$ (bold dotted lines), $T = 0.5T_F$ (bold dashed lines) and $T=T_F$ (bold solid lines). The thin lines indicate the plasmon dispersion of SLG with the same density and temperature. Here we use the parameters: $n_1 = n_2 = 10^{12}$ cm$^{-2}$ and (a) $d = 20$ A$^0$ ($k_Fd = 0.35$), (b) $d = 100$ A$^0$ ($k_Fd = 1.8$), (c) $d = 300$ A$^0$ ($k_Fd = 5.3$), and (d) $d = 500$ A$^0$ ($k_Fd = 8.9$)

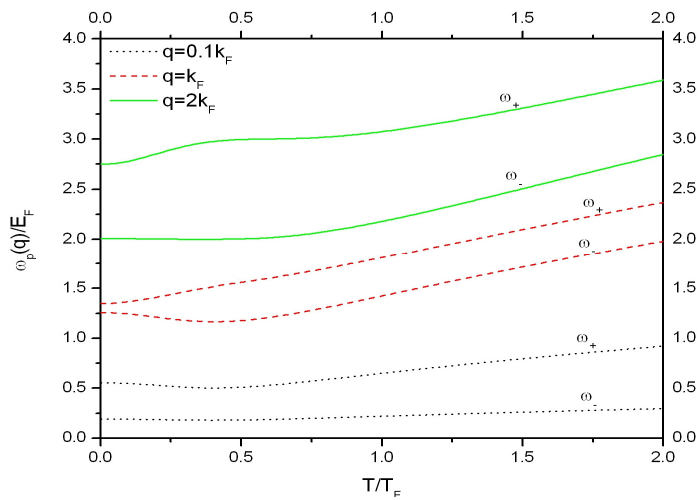

FIG. 3. (Color online) The plasmon modes of DLG at several momentums and fixed layer separation $d= 100$A$^0$

In Fig. 3, we show the temperature effect on the plasmon mode dispersion at several momentums and fixed layer separation $d=100$A$^0$. When $T < T_0$ ($T > T_0$) ($T_0 \approx 0.4T_F$) the acoustic mode $\omega_-$ decreases (increases) when the temperature $T$ increases. The Fig. 3 shows that in the case of low momentum, the optical mode $\omega_+$ decreases and then increases while it only increases in the cases of large momentum. The Fig. 3 also shows that the high temperature plasmon modes are dramatically different from the zero temperature ones especially in the case of large momentum.

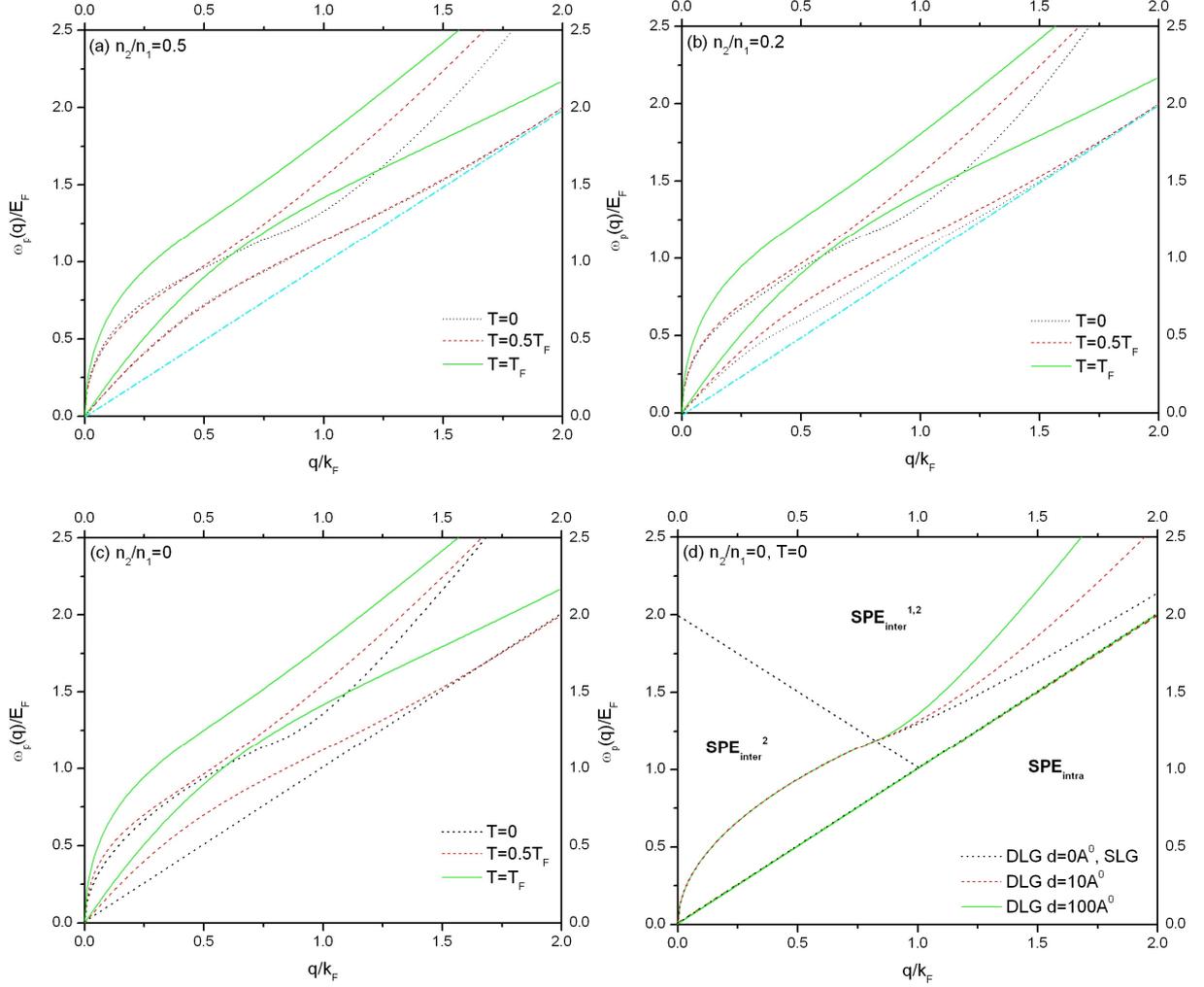

FIG. 4. (Color online) Plasmon dispersions of DLG for different temperatures and layer densities. Here we use $n_1 = 10^{12}$ cm$^{-2}$ and $d = 100$ A$^0$.

In Fig. 4 we show the calculated DLG plasmon dispersions for different layer densities. We observe that the plasmon dispersion is almost unchanged at high temperature ($T \approx T_F$). As $n_2/n_1$ decreases, the zero temperature acoustic mode $\omega_-$ approaches the boundary of the intraband single-particle excitation (SPE$_{intra}$) but it shifts, however, to higher energy when the temperature increases. In Fig. 4(d) we show the DLG plasmon modes at zero temperature for several separations in the case of $n_2/n_1 = 0$, i.e., the second layer is undoped and the first layer has a finite density $n_1 = 10^{12}$ cm$^{-2}$. It is seen from the figure that the acoustic mode $\omega_-$ is degenerate with the boundary of SPE$_{intra}$ (i.e. $\omega = v_F q$), while the optical mode $\omega_+$ is degenerate with the SLG optical mode below the SPE$_{inter}^{1,2}$. As $d \to 0$ the dispersion of $\omega_+$ becomes exactly that of SLG plasmon with the same density.

*3.2 The loss function*

In Fig. 5 we show the density plots of DLG loss function (i.e., $-\Im m[\varepsilon_{do}(q,\omega,T)^{-1}]$) in $q - \omega$ space for different separations and temperatures where the darkness represents the mode spectral strength.

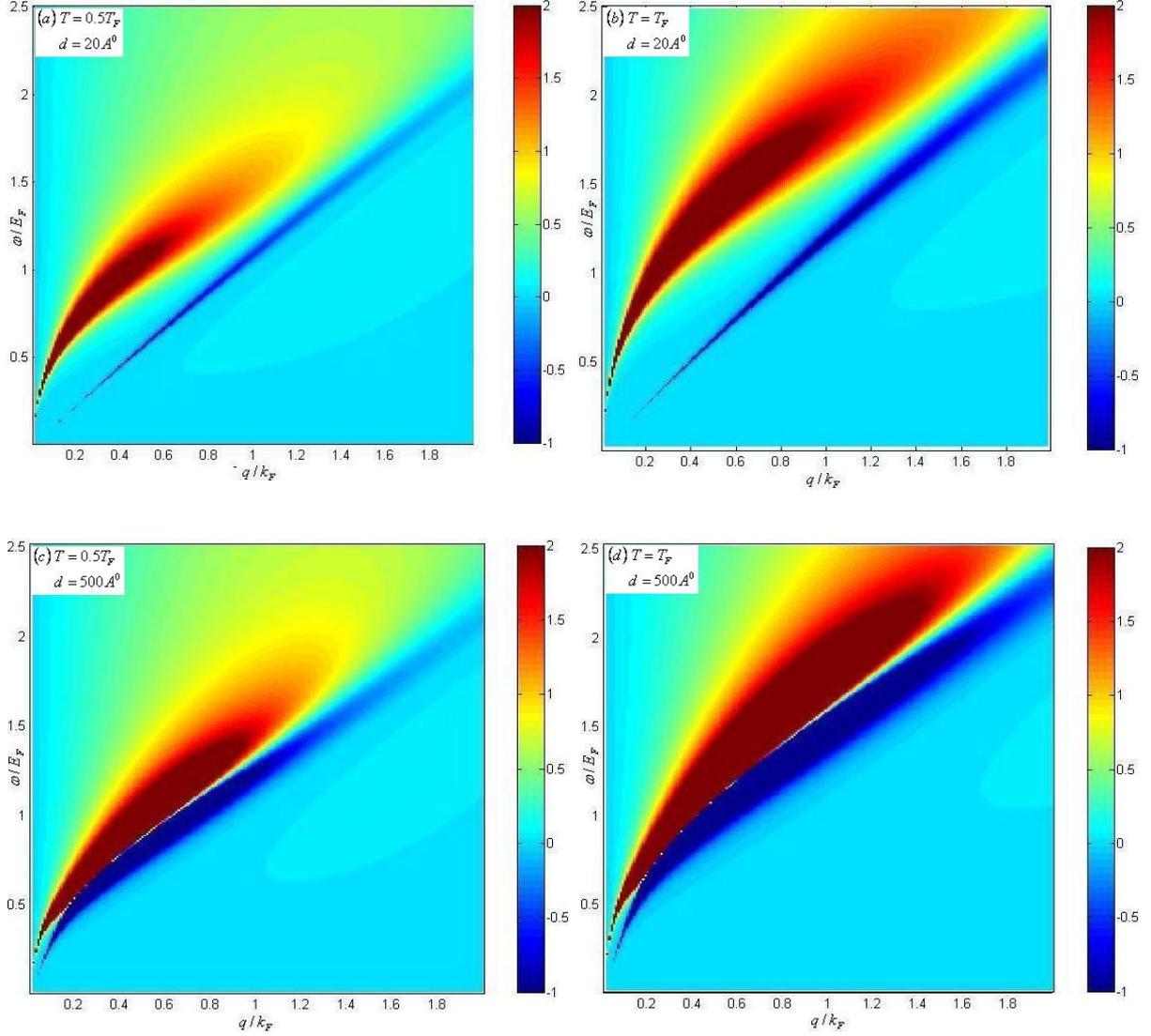

FIG. 5. (Color online) The density plot of DLG loss function in $(q,\omega)$ space for fixed densities of $n_1 = n_2 = 10^{12} \text{cm}^{-2}$, layer separations $d = 20\text{A}^0$, $500\text{A}^0$ and temperatures $T = 0.5T_F$, $T_F$.

The loss function is related to the dynamical structure factor $S(q,\omega)$, which gives a direct measure of the spectral strength of the various elementary excitations. Thus, our results can be measured in experiments such as inelastic electron and Raman-scattering spectroscopies [13, 14]. In the region below the $\omega = v_F q$ line, the value of loss function is almost zero. The acoustic mode $\omega_-$ corresponds to a broadened peak near the $\omega = v_F q$ line and the optical mode $\omega_+$ corresponds to broad peak with higher energy. As $T$ or $d$ increases, the spectral strengths of both modes $\omega_\pm$ increase.

We also calculate the loss function in the case of different layer densities. We find that the spectral strengths of both modes $\omega_\pm$ slightly increase when the density imbalance decreases (i.e., $n_2/n_1$ increases) and the high temperature spectral strength ($T \geq T_F$) is almost independent of $n_2/n_1$

In Fig. 6(a) we show the plasmon modes at $T = 0.5T_F$ for $n_2 = 0$ and $n_1 = 10^{12}\text{cm}^{-2}$. Unlike the zero temperature case shown in Fig. 4(d), the acoustic mode $\omega_-$ is degenerate with the $\omega = v_F q$ line only when $d = 0\text{A}^0$. More interestingly, even though there are no free carriers in the second layer and $d = 0$ the optical mode $\omega_+$ (the dotted line) does not become the SLG plasmon (the dot-dashed line) with the same density. In Figs. 6(b)-6(d) we show the loss function of DLG corresponding to Fig. 6(a). As the layer separation decreases, the acoustic mode $\omega_-$ approaches the $\omega = v_F q$ line and loses its spectral strength severely.

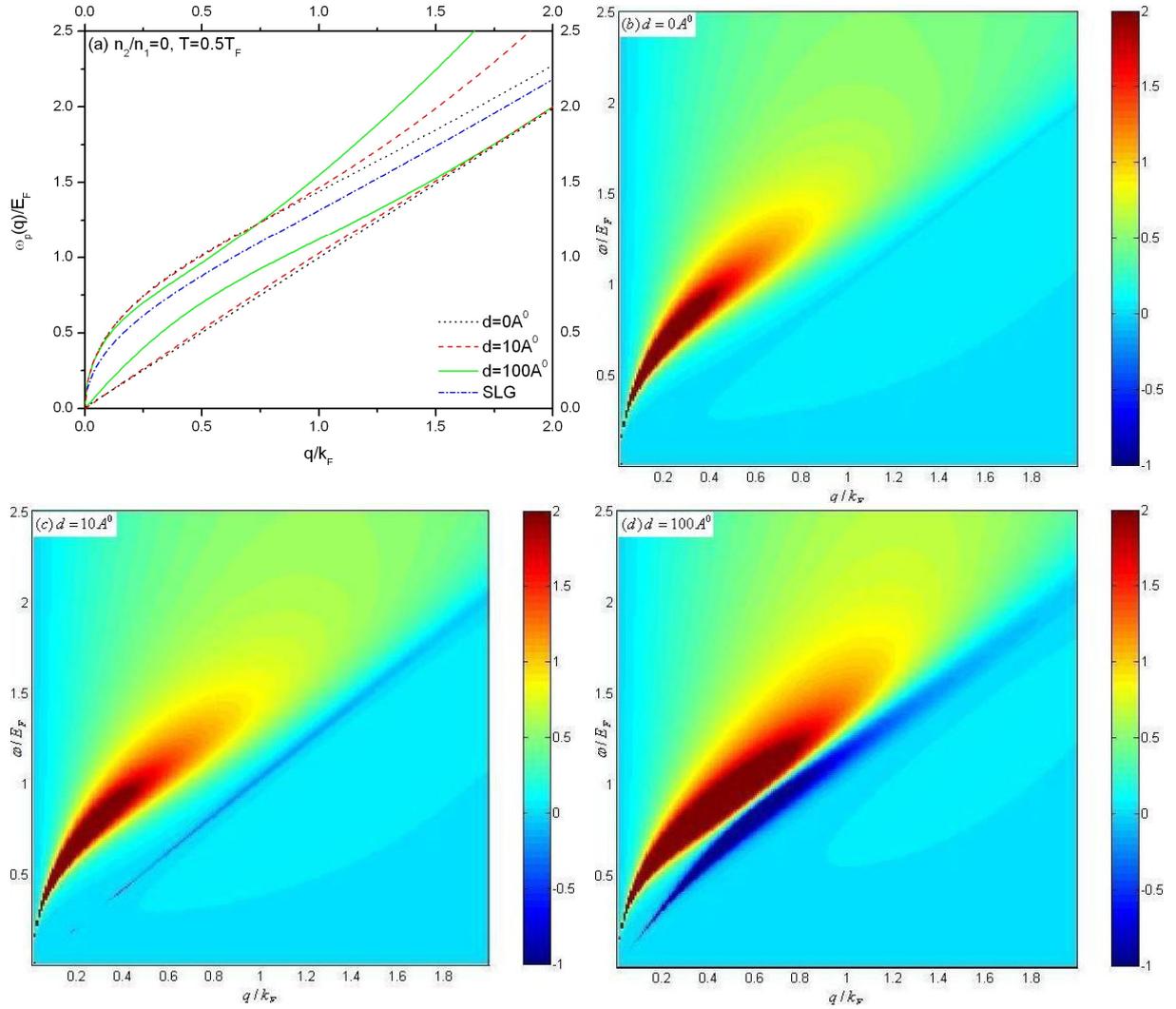

FIG. 6. (Color online) (a) Calculated plasmon mode dispersions of DLG for different layer separations for $T = 0.5T_F$, $n_2/n_1 = 0$ (i.e., $n_2 = 0$ and $n_1 = 10^{12}\text{cm}^{-2}$), and corresponding loss functions for (b) $d = 0\text{A}^0$, (c) $d = 10\text{A}^0$, and (d) $d = 100\text{A}^0$.

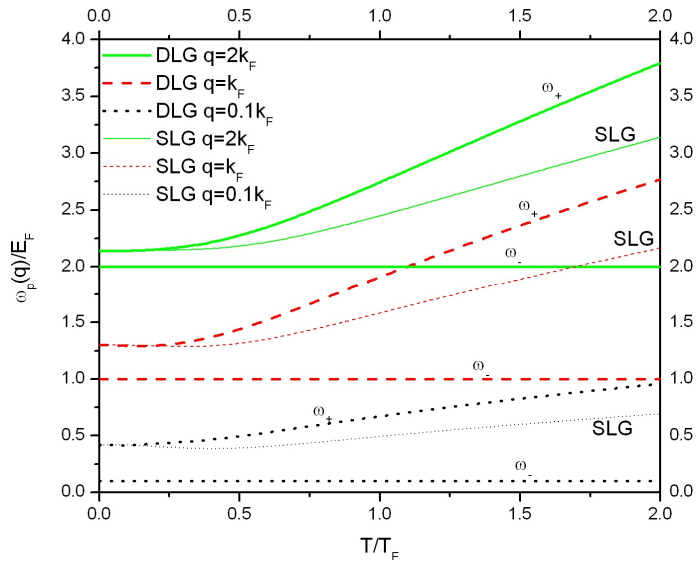

FIG. 7. (Color online) The plasmon modes of DLG at several momentums for $n_2=0$ and $d = 0\text{A}^0$

In Fig. 7, we show the temperature effect on the plasmon mode dispersion at several momentums for $n_2=0$ and $d=0A^0$. The optical modes $\omega_+$ approach the SLG plasmon modes at low temperature and shift to higher energy at high temperature.

## 4. Conclusions

In this paper, we have investigated the temperature effect on the plasmon dispersion mode and loss function of doped DLG, made of two parallel graphene monolayers with carrier densities $n_1$, $n_2$, and an interlayer separation of $d$. We have shown that unlike the zero temperature plasmon modes, the temperature acoustic mode $\omega_-$ in the case of $n_2=0$ is degenerate with the $\omega = v_F q$ line only when $d = 0 \text{A}^0$. More interestingly, even though there are no free carriers in the second layer, $n_2=0$, and $d = 0$ the temperature optical mode $\omega_+$ does not become the SLG plasmon with the same density. Our results indicate that the effect of temperature on the plasmon dispersion and damping is significant and can not be ignored in investigating many graphene properties.

*Acknowledgement-* We thank Do Hoang Son and Nguyen Thanh Son for useful discussions. This work is supported by the Vietnam's National Foundation for Science and Technology Development (NAFOSTED).